\documentclass{emulateapj}

\usepackage{amsmath}
\usepackage{epsfig}
\usepackage{amssymb}
\usepackage{graphicx}
\usepackage{color}
\usepackage{natbib}

\def\gtaprx {\lower .1ex\hbox{\rlap{\raise .6ex\hbox{\hskip .3ex
	{\ifmmode{\scriptscriptstyle >}\else
		{$\scriptscriptstyle >$}\fi}}}
	\kern -.4ex{\ifmmode{\scriptscriptstyle \sim}\else
		{$\scriptscriptstyle\sim$}\fi}}}
\def\ltaprx {\lower .1ex\hbox{\rlap{\raise .6ex\hbox{\hskip .3ex
	{\ifmmode{\scriptscriptstyle <}\else
		{$\scriptscriptstyle <$}\fi}}}
	\kern -.4ex{\ifmmode{\scriptscriptstyle \sim}\else
		{$\scriptscriptstyle\sim$}\fi}}}

\newcommand{\cutt}[1]{\textcolor{blue}{}}

\newcommand{\Ms}{{\ensuremath{M_{\odot} }}}
\newcommand{\Ls}{{\ensuremath{L_{\odot} }}}

\begin{document}

\title{On the Detection of Supermassive Primordial Stars}

\author{Marco Surace,\altaffilmark{1} Daniel J. Whalen,\altaffilmark{1} Tilman Hartwig,\altaffilmark{2,3} Erik Zackrisson,\altaffilmark{4} S.~C.~O. Glover,\altaffilmark{5} Samuel Patrick,\altaffilmark{1} Tyrone E. Woods,\altaffilmark{6,7} Alexander Heger\altaffilmark{7,8} and Lionel Haemmerl\'{e}\altaffilmark{9}}

\altaffiltext{1}{Institute of Cosmology and Gravitation, University of Portsmouth, Portsmouth PO1 3FX, UK}
\altaffiltext{2}{Department of Physics, School of Science, University of Tokyo, Bunkyo, Tokyo 113-0033, Japan}
\altaffiltext{3}{Kavli IPMU (WPI), UTIAS, The University of Tokyo, Kashiwa, Chiba 277-8583, Japan}
\altaffiltext{4}{Department of Physics and Astronomy, Uppsala University, Box 516, SE-751 20 Uppsala, Sweden}
\altaffiltext{5}{Universit\"at Heidelberg, Institut f\"ur Theoretische Astrophysik, Albert-Ueberle-Str. 2, 69120 Heidelberg, Germany}
\altaffiltext{6}{School of Physics and Astronomy, Birmingham University, Birmingham B15 2TT, United Kingdom}
\altaffiltext{7}{Monash Centre for Astrophysics, School of Physics and Astronomy, Monash University, VIC 3800, Australia}
\altaffiltext{8}{Tsung-Dao Lee Institute, Shanghai 200240, China}
\altaffiltext{9}{Observatoire de Gen\`eve, Universit\'e de Gen\`eve, chemin des Maillettes 51, CH-1290 Sauverny, Switzerland}

\begin{abstract}

The collapse of supermassive primordial stars in hot, atomically-cooled halos may have given birth to the first quasars at $z \sim$ 15 - 20.  Recent numerical simulations of these rapidly accreting stars reveal that they are cool, red hypergiants shrouded by dense envelopes of pristine atomically-cooled gas at 6,000 - 8,000 K, with luminosities $L$ $\gtrsim$ 10$^{10}$ \Ls. Could such luminous but cool objects be detected as the first stage of quasar formation in future near infrared (NIR) surveys?  We have now calculated the spectra of supermassive primordial stars in their birth envelopes with the Cloudy code. We find that some of these stars will be visible to the {\em James Webb Space Telescope} ({\em JWST}) at $z \lesssim$ 20 and that with modest gravitational lensing {\em Euclid} and the {\em Wide-Field Infrared Space Telescope} ({\em WFIRST}) could detect them out to $z \sim$ 10 - 12.  Rather than obscuring the star, its accretion envelope enhances its visibility in the NIR today by reprocessing its short-wavelength flux into photons that are just redward of the Lyman limit in the rest frame of the star.
 
\end{abstract}

\keywords{quasars: general --- black hole physics --- early universe --- dark ages, reionization, first stars --- galaxies: formation --- galaxies: high-redshift}

\maketitle

\section{Introduction}

Supermassive primordial stars (SMSs) may have been the origin of the first quasars, a few of which have now been discovered at $z >$ 7 \citep{mort11,ban18}.  These stars are thought to form in primordial halos exposed to either unusually strong Lyman-Werner (LW) UV fluxes \citep{agarw15} or highly supersonic baryon streaming motions \citep{hir17,srg17}. Either one can prevent primordial halos from forming stars until they reach masses of 10$^7$ - 10$^8$ \Ms\ and virial temperatures of $\sim$ 10$^4$ K that trigger rapid atomic cooling and catastrophic baryon collapse at central infall rates of up to $\sim$ 1 \Ms\ yr$^{-1}$ \citep{bl03,ln06,wta08,rh09b,latif13a}.

Stellar evolution models show that Population III (Pop III) stars growing at these rates can reach masses of a few 10$^5$ \Ms. Most then collapse to black holes \citep[direct collapse black holes, or DCBHs;][]{um16,tyr17,hle18} via the general relativistic (GR) instability, although a few non-accreting stars have been found to explode as highly energetic thermonuclear transients \citep{wet12a,jet13a,wet13b,wet13d,chen14b}. Pop III SMSs are currently the leading contenders for the seeds of the earliest supermassive black holes (SMBHs) because the environments of ordinary Pop III star BHs are less conducive to their rapid growth \citep{wan04,awa09,wf12,srd18}. DCBHs are born with large masses in high densities in host galaxies that can retain their fuel supply, even when it is heated by X-rays \citep{jet13}.  

\begin{figure*} 
\begin{center}
\begin{tabular}{cc}
\includegraphics[width=0.4\textwidth]{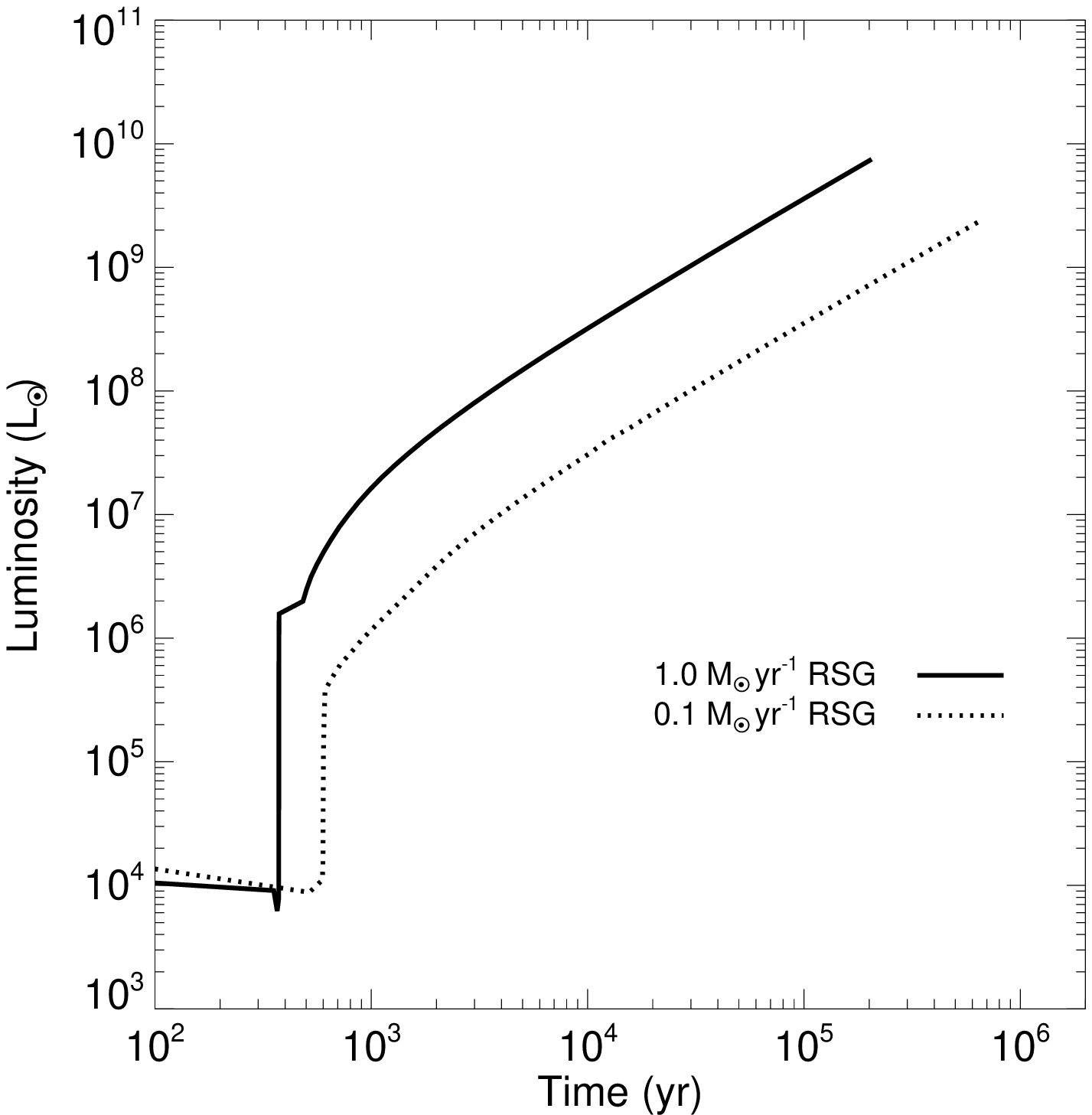}  &
\includegraphics[width=0.4\textwidth]{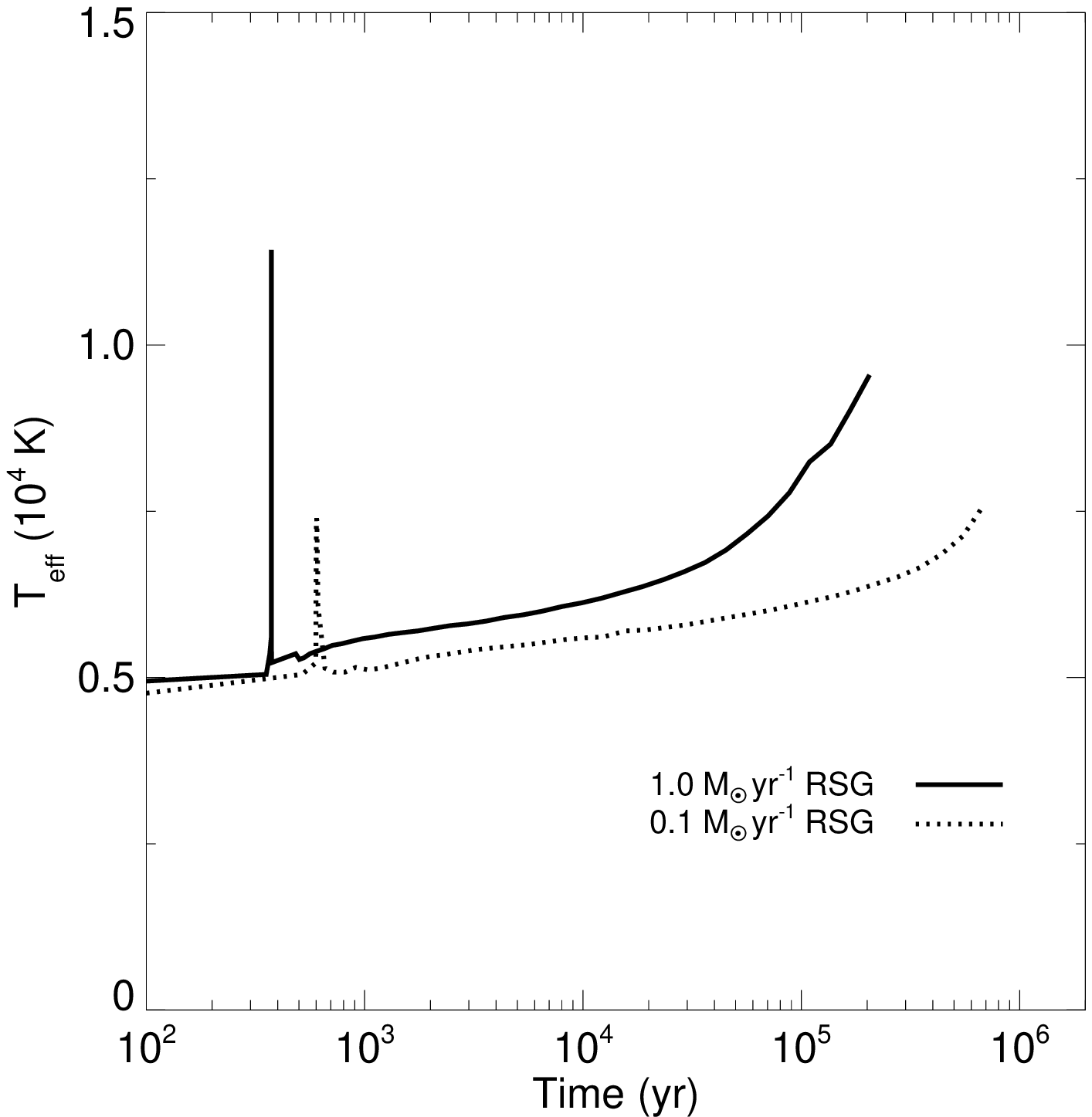}  
\end{tabular}
\end{center}
\caption{Evolution of red supergiant (RSG) stars accreting at 1.0 and 0.1 \Ms\ yr$^{-1}$ in the GENEVA stellar evolution code.  Left panel: luminosities.  Right panel: surface temperatures.}
\vspace{0.1in}
\label{fig:evol} 
\end{figure*}

What are the prospects for detecting SMSs at high redshifts?  \citet{til18a} found that the relics of such stars would be uniquely identifiable with the gravitational wave detector LISA at $z >$ 15 if they form in binaries. \citet{jlj12a} examined some spectral features of hot, blue, rapidly-accreting SMSs and found that they would be characterized by strong Balmer emission and the conspicuous absence of Ly$\alpha$ lines due to absorption by their envelopes. The source of this flux was not the star but its hypercompact H II region, whose ionizing radiation was trapped close to its surface by the density and ram pressure of the inflow \citep[which was also found to be true in cosmological simulations of highly resolved atomically cooled halos;][]{bec18}.  \citet{freese10}, \citet{z10b} and \citet{z10a} calculated the spectral signatures of hot, blue Pop III 'dark stars', supermassive primordial stars powered by the self-annihilation of weakly-interacting dark matter rather than by nuclear fusion. They found that such objects might be visible even to $8-10\,\mathrm{m}$ telescopes on the ground today, primarily because of their high surface temperatures (20,000 - 30,000 K), larger masses (up to 10$^7$ \Ms) and longer lives (up to 10$^7$ yr).  

But several studies have now shown that rapidly accreting Pop III stars generally evolve as cool, red hypergiants along the Hayashi limit with surface temperatures of 5,000 - 10,000 K \citep{hos13}.  \citet{hle17} found that such stars can reach luminosities $\gtrsim 10^{10}$ \Ls\ that could in principle be visible to {\em JWST} \citep{jwst2}, {\em Euclid}, {\em WFIRST} and extemely large telescopes (ELTs) on the ground. However, they are shrouded by dense accretion flows that reprocess radiation from the star, perhaps suppressing its flux in the NIR today.  Detecting SMSs at high redshift would capture primordial quasars at the earliest stages of their development and reveal one of their channels of formation. Here, we calculate NIR luminosities for Pop III SMSs in their accretion envelopes whose structures are taken from a high-resolution cosmological simulation. We describe our models in Section~2, calculate SMS spectra and NIR magnitudes in Section~3 and conclude in Section~4.

\section{Numerical Method}

\begin{figure*} 
\begin{center}
\begin{tabular}{cc}
\epsfig{file=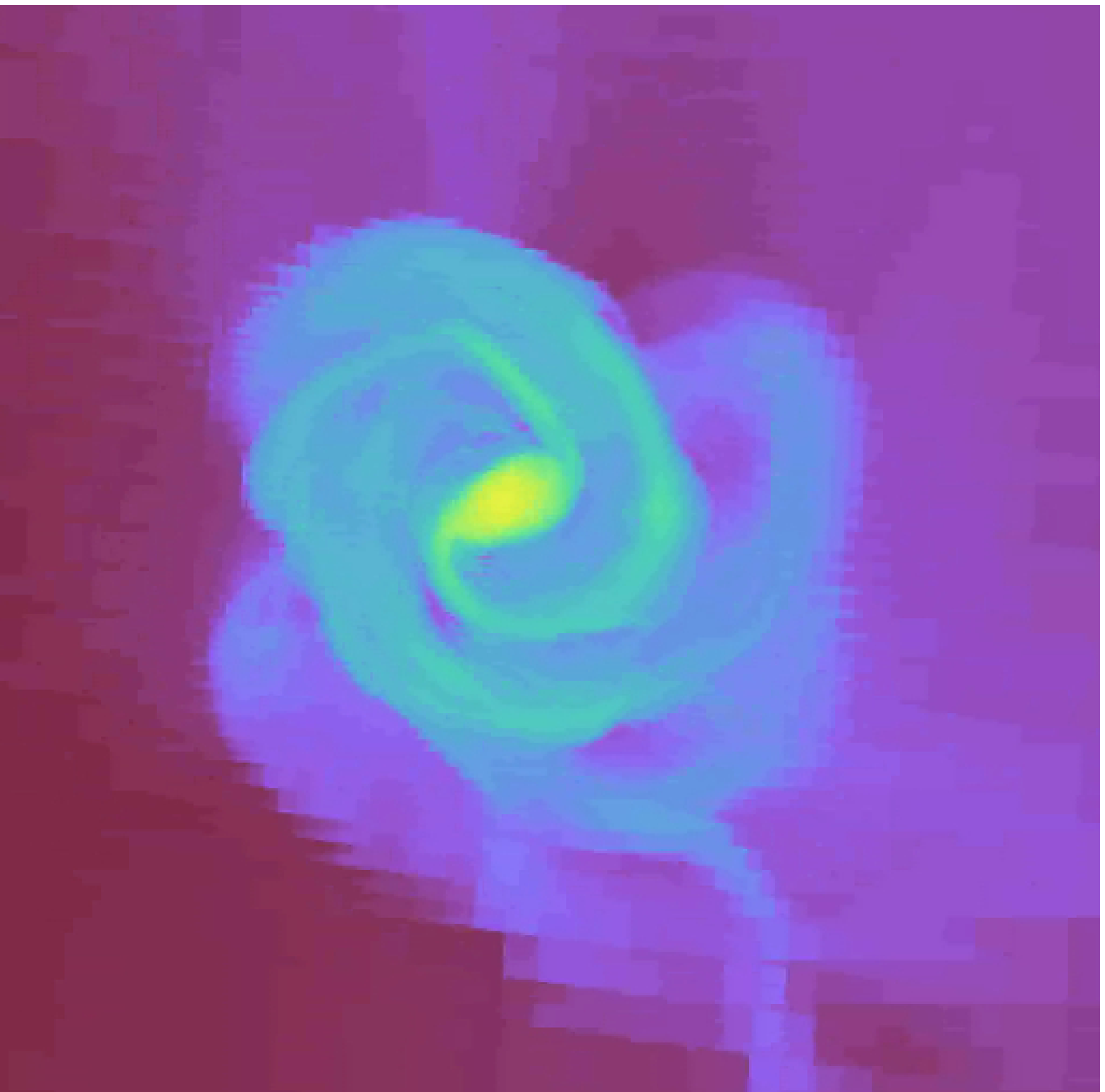,width=0.335\linewidth,clip=}  &
\epsfig{file=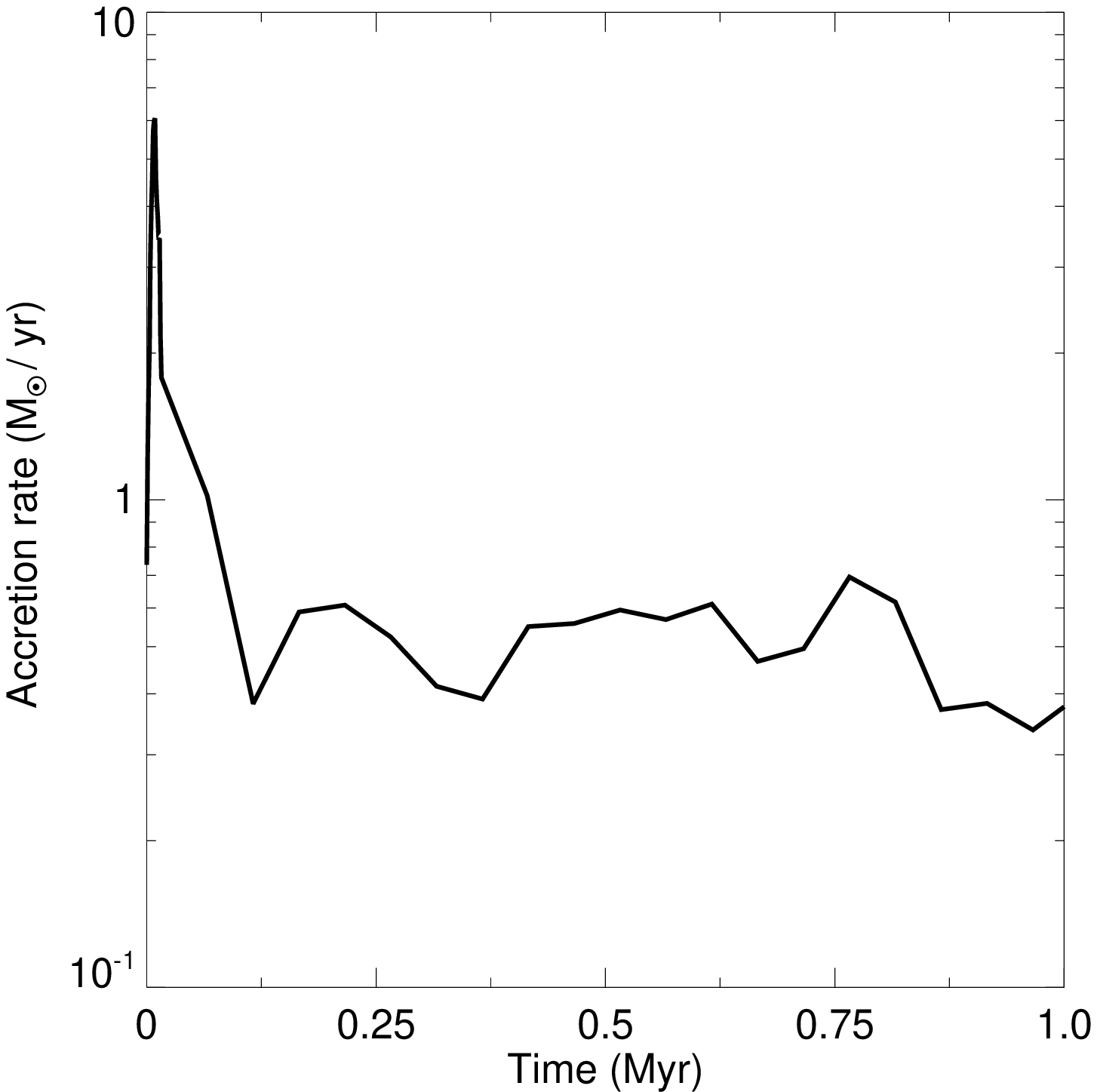,width=0.35\linewidth,clip=}  \\
\epsfig{file=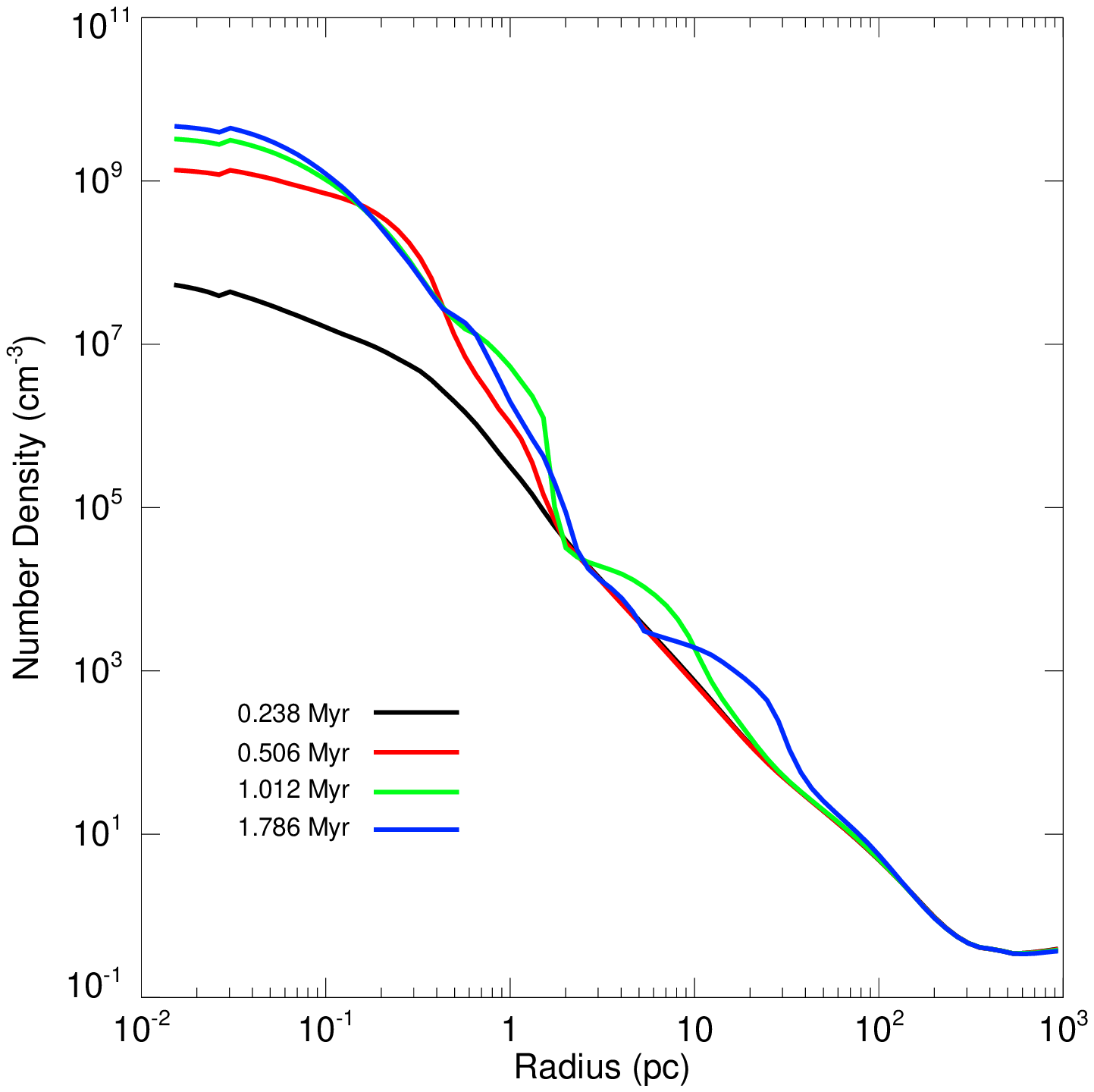,width=0.35\linewidth,clip=}  &
\epsfig{file=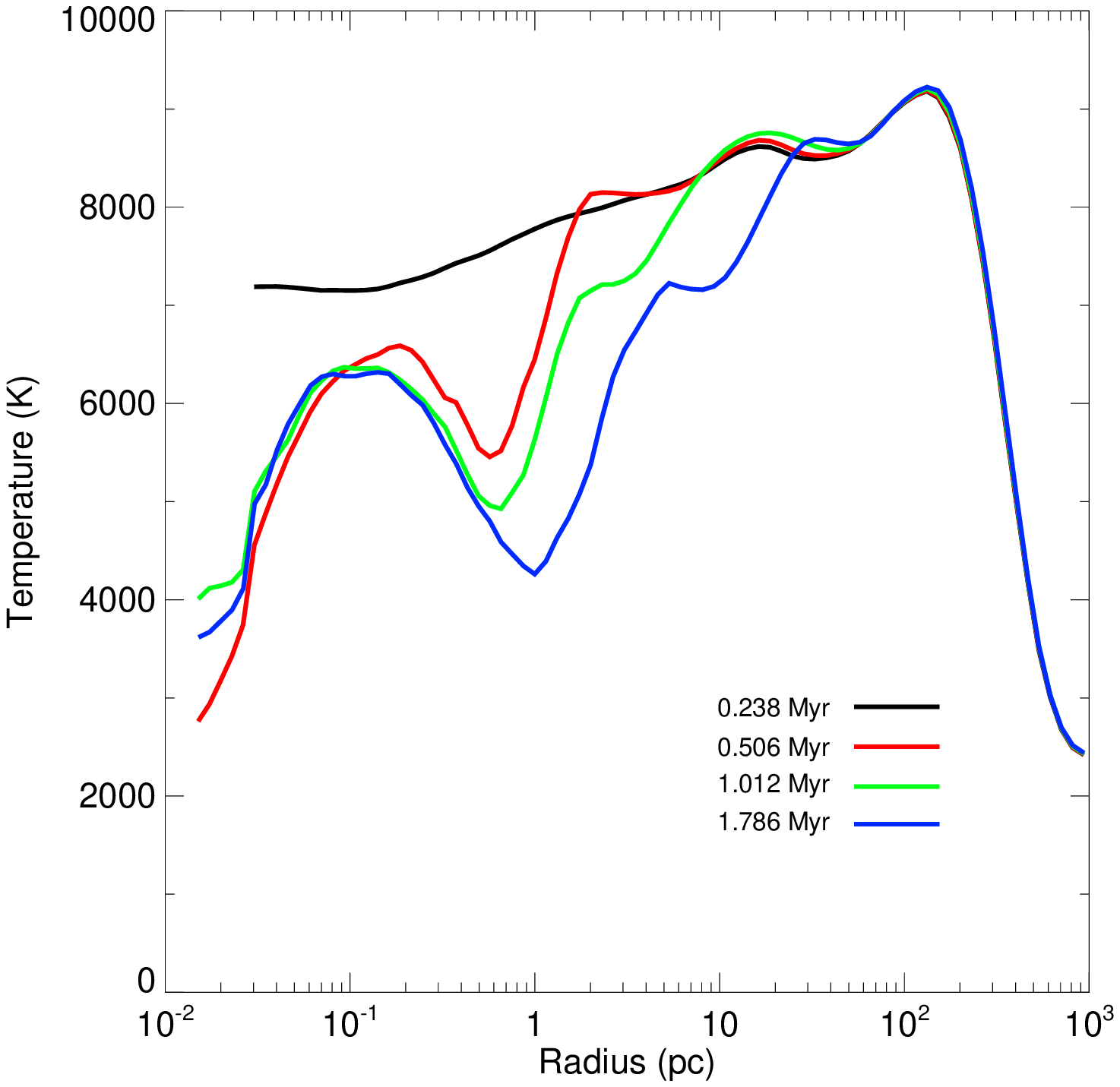,width=0.35\linewidth,clip=}  
\end{tabular}
\end{center}
\caption{Top left: accretion disk at 0.625 Myr. Top right: central accretion rates. Bottom left: spherically-averaged gas densities in the halo.  Bottom right: spherically-averaged temperatures.}
\vspace{0.1in}
\label{fig:halo} 
\end{figure*}

Rest frame spectra for the star in its envelope are calculated with Cloudy \citep{cloudy17} with envelope profiles taken from a simulation of the collapse of an atomically-cooled halo done with Enzo \citep{enzo}. The spectra are then redshifted, dimmed, and convolved with a variety of filter functions to obtain AB magnitudes in the NIR as a function of SMS redshift. We consider stars accreting at 0.1 and 1.0 \Ms\ yr$^{-1}$ whose properties are listed in Tables A3 and A2 of \citet{hle17}, respectively.  Bolometric luminosities, $L_{\mathrm{bol}}$, and effective temperatures, $T_{\mathrm{eff}}$, for both stars are shown in Figure~\ref{fig:evol}. 

\subsection{Enzo Envelope Model}

The halo in which the SMS is assumed to form was evolved in a 1.5 $h^{-1}$ Mpc box in Enzo from $z =$ 200 down to $z =$ 17.8, when it reaches a mass of 2.7 $\times$ 10$^7$ \Ms\ and begins to atomically cool and collapse.  It is centered in three nested grids for an initial effective resolution of 2048$^3$, and we allow up to 15 levels of refinement for a maximum resolution of 0.014 pc. The grid is initialized at $z =$ 200 with cosmological parameters taken from the second-year \textit{Planck} release: $\Omega_{\mathrm{M}}=0.308$, $\Omega_{\Lambda}=0.691$, $\Omega_{\mathrm{b}}=0.0223$, $h=0.677$, $\sigma_8=0.816$, and $n=0.968$ \citep{planck2}. To approximate the presence of a strong LW background, we evolve the halo without H$_2$, just H, H$^+$, e$^-$, He, He$^+$ and He$^{++}$ \citep{grackle}. Cooling by collisional ionization and excitation of H and He, bremsstrahlung, and inverse Compton scattering are all included in the energy equation.  

As shown in the upper left panel of Figure~\ref{fig:halo}, a large atomically cooled disk forms at the center of the halo that is $\sim$ 2 pc in diameter and at 4,000 - 6,000 K at 0.625 Myr after the onset of collapse.  A bar instability in the disk efficiently transports angular momentum out of its center, producing the large sustained accretion rates onto the star shown in the upper right panel of Figure~\ref{fig:halo}. After a brief burst due to initial collapse and the formation of the disk, infall proceeds at rates of 0.4 - 0.6 \Ms\ yr$^{-1}$. Spherically averaged density and temperature profiles of the halo are shown in the bottom two panels of Figure~\ref{fig:halo} at 0.238 Myr, 0.506 Myr, 1.012 Myr and 1.786 Myr.

\subsection{Cloudy Spectra}

We treat both stars as blackbodies (BBs) because they are relatively cool and have no absorption lines due to metals. Cloudy fits BB spectra to each star from $L_{\mathrm{bol}}$ and $T_{\mathrm{eff}}$, which we take to be 1.26 $\times$ 10$^9$ \Ls\ and 6653 K for the 0.1 \Ms\ yr$^{-1}$ star and 3.92 $\times$ 10$^9$ \Ls\ and 8241 K for the 1.0 \Ms\ yr$^{-1}$ star. These values correspond to 3.49 $\times$ 10$^5$ yr and 1.089 $\times$ 10$^5$ yr for the two stars, about halfway through their respective lifetimes.  Ideally, one would surround the star with the accretion envelope that created it in a cosmological simulation for self-consistency. But stellar evolution models of Pop III SMSs in time-dependent cosmological flows are not yet available, so we instead use density and temperature profiles from the Enzo simulation at 1.786 Myr as the envelope of each star. This choice is reasonable because the accretion rates associated with these profiles are intermediate to those in which our stars were evolved. 

The density and temperature profiles of the envelope that are input to Cloudy are tabulated in 70 bins that are uniformly partitioned in log radius, with inner and outer boundaries at 0.015 pc and 927 pc. Cloudy solves the equations of radiative transfer, statistical and thermal equilibrium, ionization and recombination, and heating and cooling to calculate the excitation and ionization state of the gas surrounding the star and compute its emergent spectrum. The temperatures of the gas falling onto the star are set by the virialization of cosmic flows well above it, not by radiation from the star. Since they determine to what degree the envelope is collisionally excited, and thus how it reprocesses photons from the star, we required Cloudy to use the temperatures Enzo calculates for the envelope instead of inferring them from the spectrum of the SMS because they would have been too low.

\section{Observing Supermassive Stars}

\subsection{SMS Spectra}

\begin{figure} 
\plotone{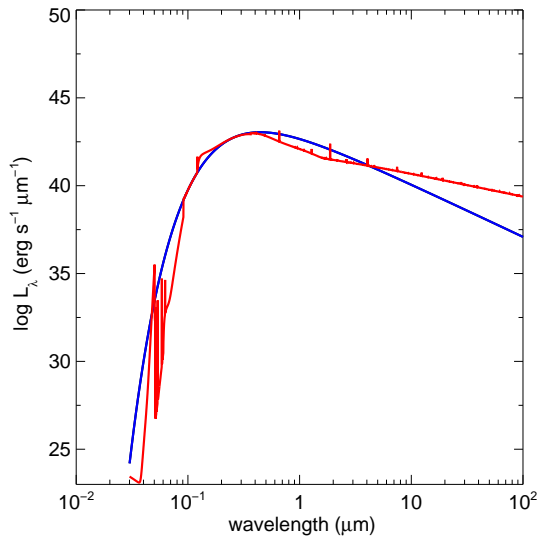}
\caption{1.0 \Ms\ yr$^{-1}$ SMS spectra at 100,000 yr.  Blue: spectrum of the star itself; red: spectrum after reprocessing by the envelope.}
\vspace{0.1in}
\label{fig:spec} 
\end{figure}

We show spectra for the 1.0 \Ms\ yr$^{-1}$ star at 1.089 $\times$ 10$^5$ yr before and after attenuation by its envelope in Figure~\ref{fig:spec}.  As expected, the stellar spectrum peaks at 0.352 $\mu$m and absorption by the envelope at the Lyman limit of H is clearly visible at 0.0912 $\mu$m.  The continuum absorption below 0.0912 $\mu$m is punctuated by several prominent He emission lines.  There is a Ly-$\alpha$ emission line at 0.1216 $\mu$m and strong H$\alpha$ and Paschen series lines are visible at 0.656 $\mu$m, 1.28 $\mu$m and 1.88 $\mu$m. There is continuum absorption half a decade in wavelength above and below 1.65 $\mu$m due to H$^-$ bound-bound and bound-free opacity, respectively.  

Photons from the star that are blueward of the Lyman limit are reprocessed by its envelope into the Ly$\alpha$ and two-photon continuum emission visible at 0.1216 - 0.16 $\mu$m.  This latter flux is greater than that emitted by the star itself and can enhance its visibility in the NIR today.  The effect varies with $T_{\mathrm{eff}}$ and source redshift but is at most 0.5 - 1 AB mag.  The Ly$\alpha$ will not aid in the detection of the star because it will be scattered into a halo of low surface brightness in the neutral IGM.  We note that at the velocities and densities of the infall onto the surface of the star, the luminosity of its accretion shock is at most $\sim$ 10$^4$ \Ls\ and does not produce a significant contribution to the visibility of the SMS.

\subsection{NIR Magnitudes}

\begin{figure*} 
\begin{center}
\begin{tabular}{cc}
\epsfig{file=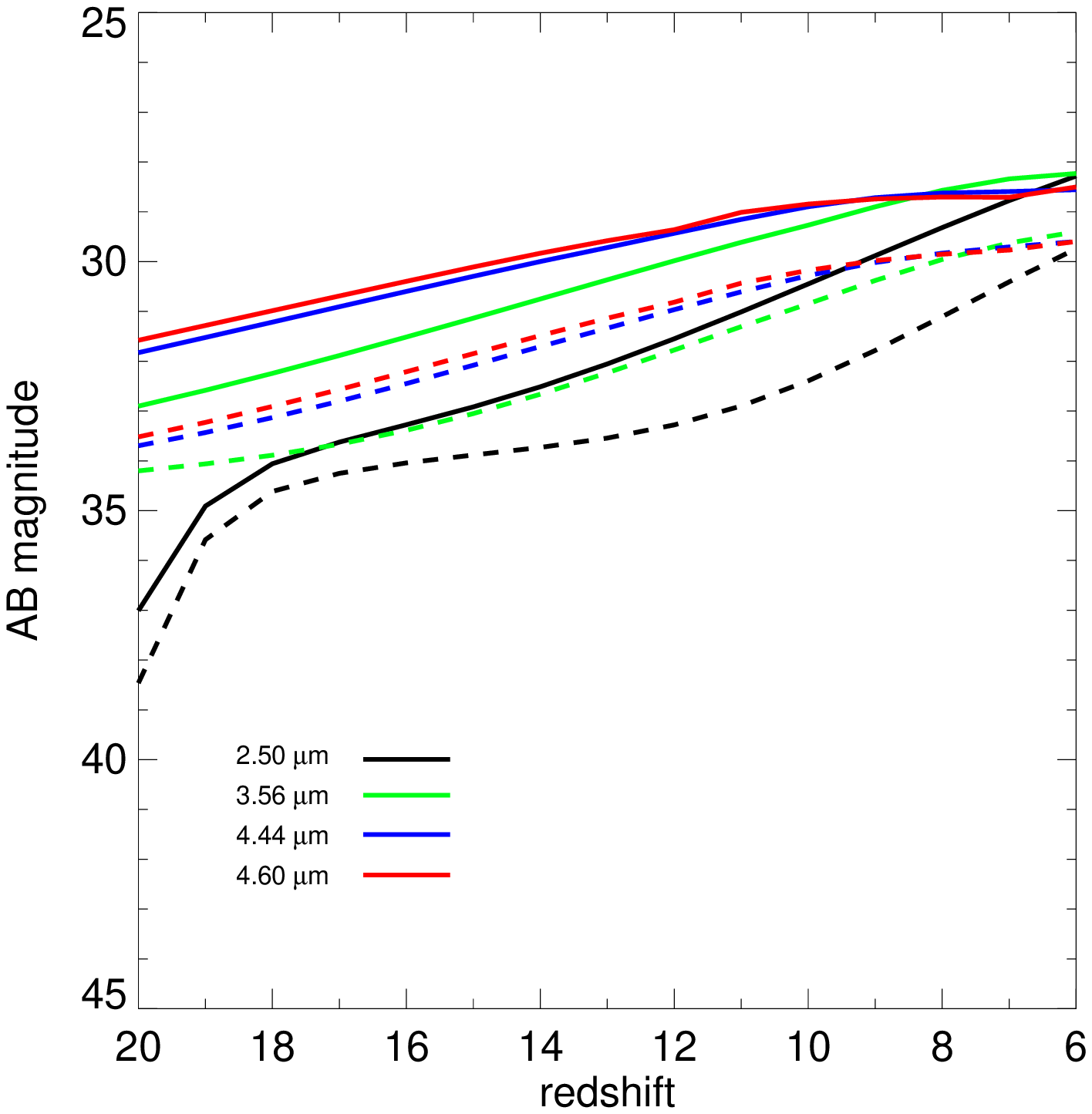,width=0.35\linewidth,clip=}  &
\epsfig{file=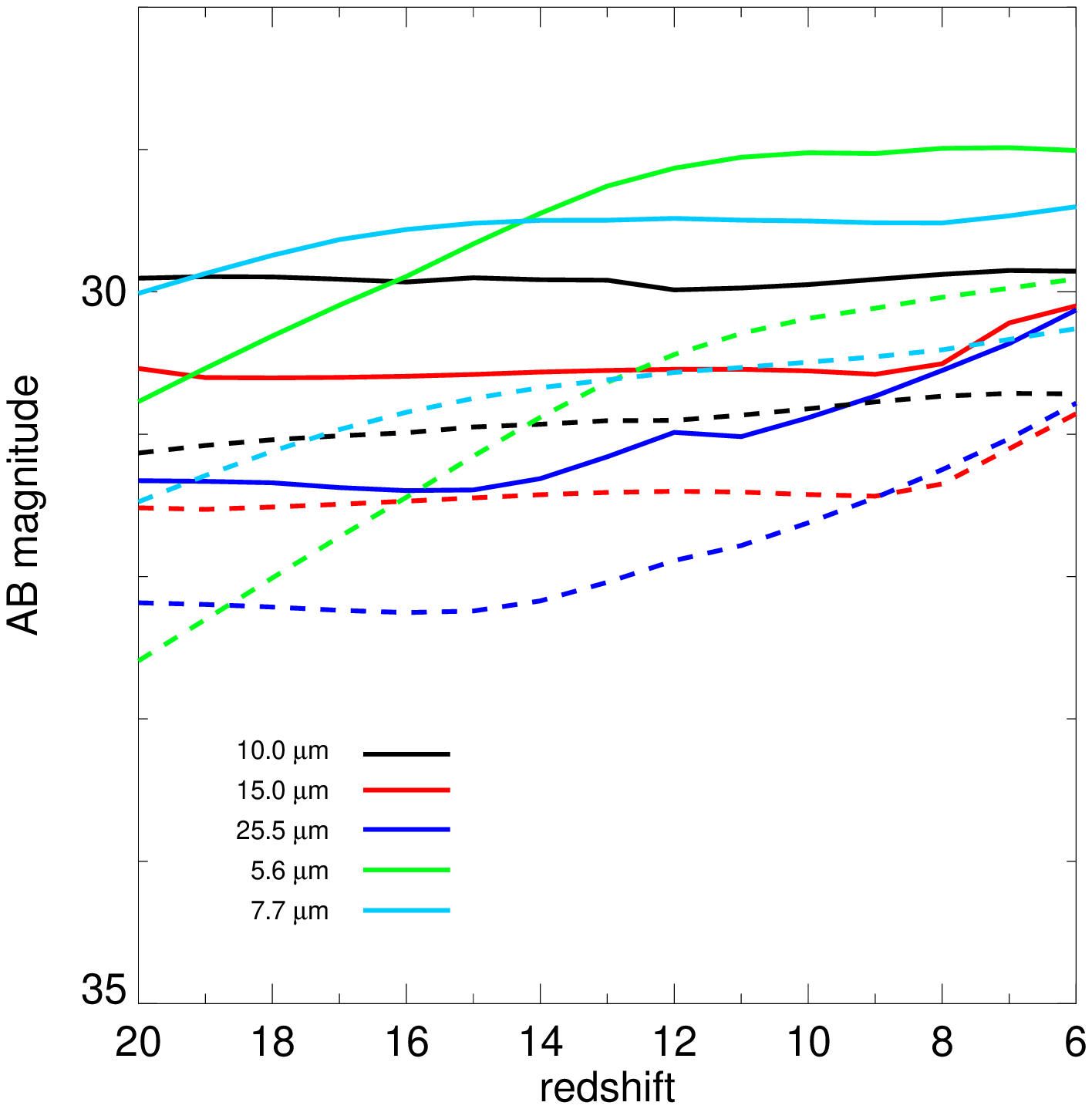,width=0.35\linewidth,clip=}  \\
\epsfig{file=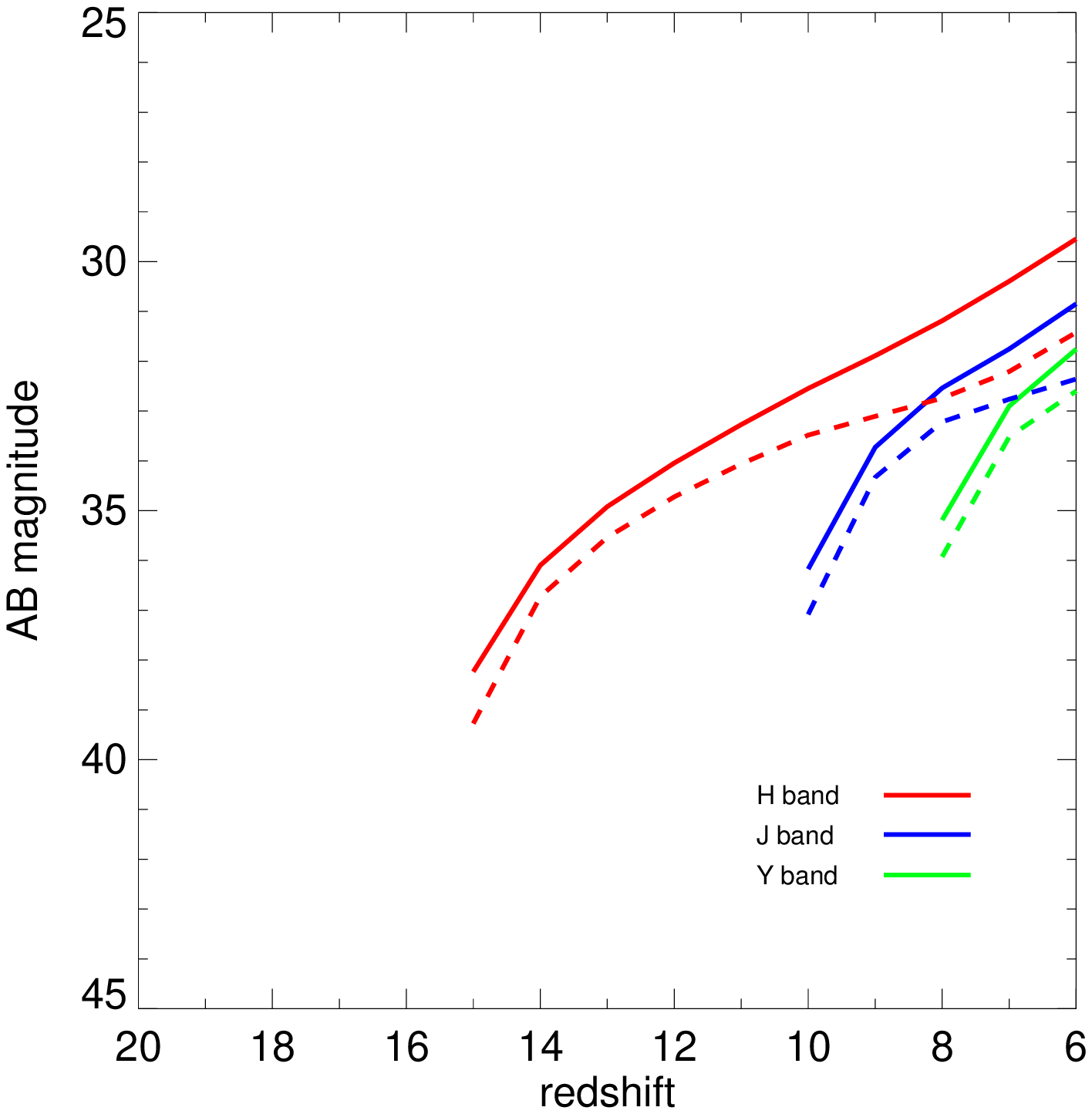,width=0.35\linewidth,clip=}  &
\epsfig{file=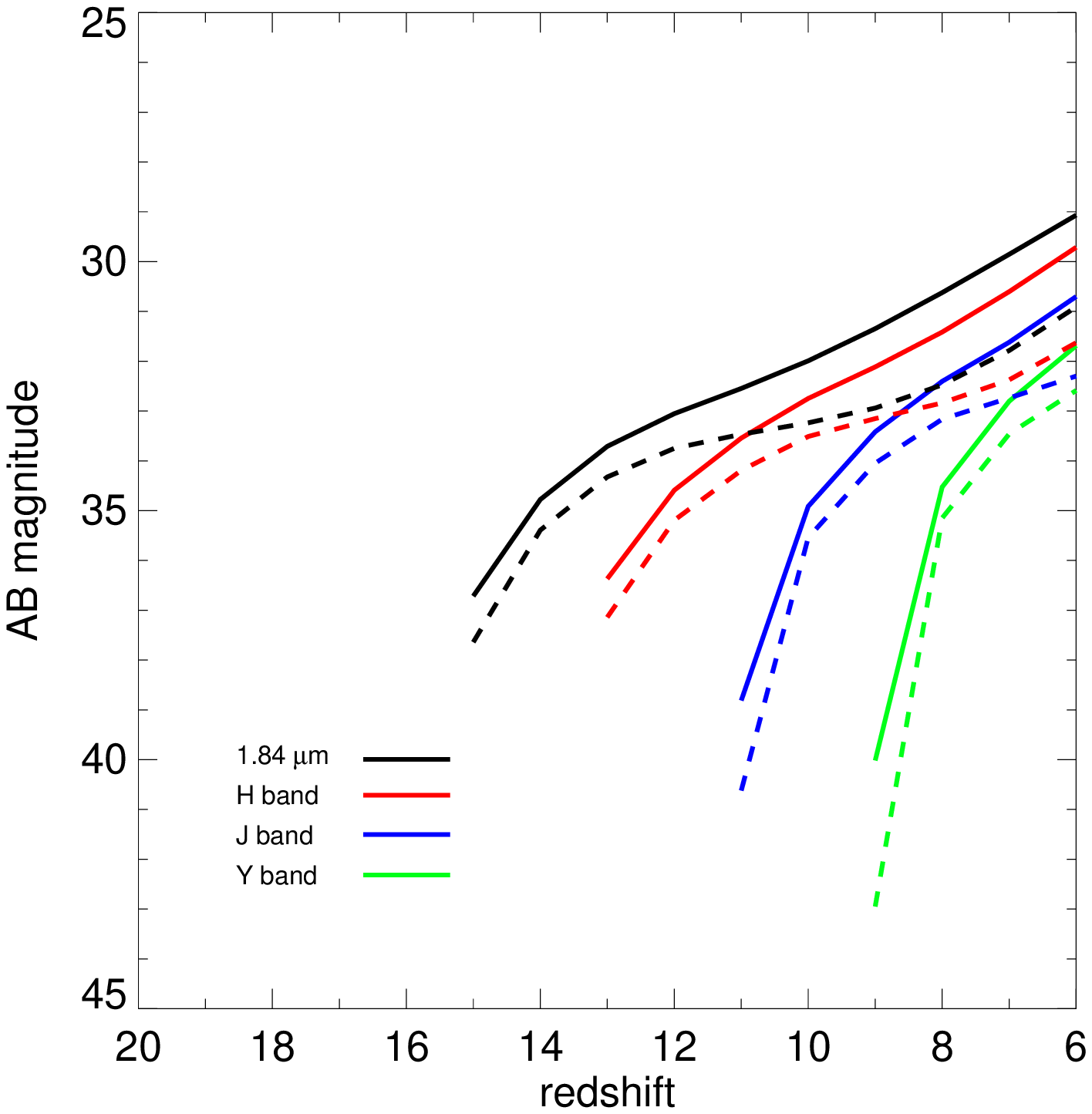,width=0.35\linewidth,clip=}  
\end{tabular}
\end{center}
\caption{NIR AB magnitudes for the 1.0 \Ms\ yr$^{-1}$ (solid) and 0.1 \Ms\ yr$^{-1}$ (dashed) SMSs in {\em JWST}, {\em Euclid} and {\em WFIRST} bands. Top left: {\em JWST} NIRCam bands.  Top right: {\em JWST} MIRI bands.  Bottom left: {\em Euclid}. Bottom right: {\em WFIRST}.}
\vspace{0.1in}
\label{fig:ABmag} 
\end{figure*}

We show AB magnitudes for both stars in {\em JWST} NIRCam bands at 2.5 - 4.6 $\mu$m in the top right panel of Figure \ref{fig:ABmag}.  The 1.0 \Ms\ yr$^{-1}$ SMS is consistently 1 - 2 magnitudes brighter than the 0.1 \Ms\ yr$^{-1}$ SMS except at high redshifts at 2.50 $\mu$m,  where both luminosities abruptly fall off because of absorption of flux blueward of Ly$\alpha$ in the source frame of the star by the neutral intergalactic medium (IGM) at $z \gtrsim$ 6. At $z \sim 6 - 8$ the two stars are brightest in the 3.56 $\mu$m filter but at $z >$ 10 they are brighter in the 4.44 $\mu$m and 4.60 $\mu$m bands, with magnitudes that vary from 28.5 - 31.5 at $z =$ 6 - 20 for the 1.0 \Ms\ yr$^{-1}$ SMS and 29.5 - 33.5 for the 0.1 \Ms\ yr$^{-1}$ SMS.  

SMS magnitudes are much more uniform in redshift in the mid infrared, as we show for several {\em JWST} MIRI bands in the top right panel of Figure~\ref{fig:ABmag}.  They exhibit the greatest variation at 5.6 $\mu$m, which is closest to the NIR, but largely level off at 7.7 - 25.5 $\mu$m.  This behavior is primarily due to the flattening of the spectrum at wavelengths above 1.5 $\mu$m in the source frame due to reprocessing of flux from the star by its envelope.  The two stars are brightest from $z =$ 6 - 20 at 5.6 - 10.0 $\mu$m, with magnitudes $\lesssim$ 31 and could therefore provide important additional spectral confirmation of SMS candidates in NIRCam.

We show SMS magnitudes for {\em Euclid} and {\em WFIRST} in the lower two panels of Figure~\ref{fig:ABmag}.  Absorption by the neutral IGM at $z \gtrsim$ 6 quenches Y, J and H band fluxes at $z \gtrsim$ 7, 10 and 14, respectively, limiting detections of these stars to these redshifts in these filters.  Magnitudes for the 1.0 \Ms\ yr$^{-1}$ star vary from 29.5 - 31.8 in {\em Euclid} and 29 - 32.5 in {\em WFIRST} at $z = 6$.  For the 0.1 \Ms\ yr$^{-1}$ star, they vary from 31.5 - 32.5 in {\em Euclid} and from 31 - 32.5 in {\em WFIRST} at the same redshift.  They drop off more rapidly with redshift than in the NIRCam bands because spectral luminosities in the source frame fall with decreasing wavelength below $\sim$ 0.3 $\mu$m. 

\subsection{SMS Formation / Detection Rates}
\label{sect:rates}

Since the lifetime of an SMS is much smaller than the Hubble time, even at the high redshifts at which it is likely to form, the number of SMSs per unit redshift per unit solid angle at a redshift $z$ can be written as
\begin{equation}
\frac{{\rm d}N}{{\rm d}z{\rm d}\Omega} = \dot{n}_{\rm SMS} \, t_{\rm SMS} \, r^{2} \frac{{\rm d}r}{{\rm d}z},
\end{equation}
where $\dot{n}_{\rm SMS}$ is the SMS formation rate per unit comoving volume, $t_{\rm SMS}$ is the characteristic lifetime of an SMS, and $r(z)$ is the comoving distance to redshift $z$, 
\begin{equation}
r(z) = \frac{c}{H_{0}} \int_{0}^{z} \frac{{\rm d}z^{\prime}}{\sqrt{\Omega_{\rm m} (1 + z^{\prime})^{3} + \Omega_{\Lambda}}}.
\end{equation}
Unfortunately, $\dot{n}_{\rm SMS}$ is poorly constrained, with theoretical models predicting number densities that vary by up to eight orders of magnitude \citep[see, e.g., the recent review of][]{titans}. These models also predict different evolutions in $\dot{n}_{\rm SMS}$ with  redshift. \citet{hab16} predict a steady increase in the comoving number density of SMSs with decreasing $z$ while \citet{rosa17} predict that most form in the narrow range $z \sim 16$--18. 
We therefore consider two toy models that bracket this range of $\dot{n}_{\rm SMS}$. 

In the first, our ``optimistic'' model, we assume that most SMSs form at $z \sim 10$--12 and that the final comoving number density is around $10^{-1} \, {\rm Mpc^{-3}}$, as in the \citet{hab16} model with a low value for $J_{\rm crit}$. In the other, our ``pessimistic'' model, we assume that most SMSs form at redshifts $z \sim 16$--18, as in \citet{rosa17}, with a final comoving number density of around $10^{-8} \, {\rm Mpc^{-3}}$. The optimistic model yields approximately $4 \times 10^{7}$ potentially observable SMSs per steradian per unit redshift, or around 30 per NIRCam field of view. On the other hand, the pessimistic model predicts only $\sim 10$ SMSs per steradian per unit redshift, meaning that any given NIRCam pointing with the appropriate limiting magnitude would have a probability of only around $10^{-5}$ of detecting an SMS.

The chances of detecting an SMS are highly dependent on the model assumed for their formation. However, since some models predict number counts high enough for one or more SMSs to be found in any sufficiently deep NIRCam image, {\em JWST} will begin to place observational constraints on these models, even if it cannot rule out extreme ones such as our pessimistic model.  We note that no SMSs have been found in the {\em Hubble} Ultra Deep Field to date because of its AB mag limit of 29 at 1.38 $\mu$m in the H band, which is well below that expected of either star even at $z \sim$ 6.

\section{Conclusion}

At NIRCam AB magnitude limits of 31.5 {\em JWST} could detect the 1.0 \Ms\ yr$^{-1}$ SMS at $z \lesssim$ 20 and the 0.1 \Ms\ yr$^{-1}$ SMS at $z \lesssim$ 13.  But the prospects for discovering such stars would be better if they could also be found by {\em Euclid} and {\em WFIRST} because their wide fields would enclose far more of them at high redshifts.  Once flagged, SMS candidates could then be studied with {\em JWST} in greater detail. However, as shown in Figure \ref{fig:ABmag}, the H band magnitudes of both stars at $z =$ 6 - 20 are above current {\em Euclid} and {\em WFIRST} detection limits (26 and 28, respectively).  

But this does not mean {\em Euclid} and {\em WFIRST} will not find these stars because only modest gravitational lensing is required to boost their fluxes above their detection limits.  The fields of view of both missions will enclose thousands of galaxy clusters and massive galaxies, and at $z \sim$ 6 - 10 magnification factors of only 10 - 100 would be required to reveal either star.  It is likely that a sufficient fraction of their survey areas will be lensed to such factors \citep{ryd18a}.  Even higher magnifications may be possible in future surveys of individual cluster lenses by {\em JWST} but at the cost of smaller lensing volumes \citep{wet13c,wind18}.

In our Enzo and Cloudy calculations we have neglected the effect of radiation pressure due to flux from the star on the flows that create it.  Modeling these effects in cosmological simulations is challenging in part because they must resolve photospheres on very small scales that preclude the codes from evolving them for long times.  \citet{aaron17} post processed simulations of highly resolved atomically cooling halos with Ly$\alpha$ photon transport and found it could exert mechanical feedback on flows in the vicinity of the star.  Radiation hydrodynamical simulations by \citet{Luo18} and \citet{ard18} that neglect resonant Ly$\alpha$ scattering found that radiation from the protostar in its early stages did not significantly alter flows in its vicinity but did suppress fragmentation, thus promoting the rapid growth of a single supermassive object.  In principle, radiation from the SMS could blow out gas and partially expose it to the IGM, but this will have little effect on the AB magnitudes of the star today because all that would be lost is the mild enhancement of UV flux redward of the Lyman limit by the envelope discussed in Section 3.1.

While we have only considered cool red supergiant stars, hotter SMSs could be easier to detect because they would produce more flux in the NIR today.  Low accretion rates \citep[$\lesssim$ 0.005 \Ms\ yr$^{-1}$;][]{hle17} or clumpy accretion due to fragmentation \citep{sak15} or turbulence in the disk can produce such stars.  However, blue SMS spectra require corrections due to absorption by their atmospheres before the flux that exits the accretion envelope can be calculated.  SMSs could also be found at higher redshifts if they exhibit pulsations that temporarily boost their fluxes above the detection limits of the wide-field surveys. Although current stellar evolution codes use implicit solvers and large time steps that do not resolve such oscillations, they can cause the star to periodically brighten and dim by an order of magnitude on timescales of a few weeks in the rest frame. Such variations might also facilitate their detection because their regularity would differentiate them from dusty, red high-$z$ quasars or low-$z$ impostors such as exoplanets. Periodic dimming and brightening could also flag these objects as high-$z$ SMSs in transient surveys proposed for {\em JWST} such as FLARE \citep{flare}. 

DCBH birth may be the next stage of primordial quasar evolution, and a number of studies have considered their prospects for detection in future NIR surveys.  These are also deeply imbedded objects in dense, atomically-cooled flows and radiative transfer techniques similar to those we have used here are required to model their spectra.  One-dimensional radiation hydrodynamics simulations of DCBH emission post processed with Cloudy have shown that they could be detected by {\em JWST} out to $z \sim$ 20 \citep{bec15,nat17}.  We are now post processing radiation hydrodynamical simulations of the H II region of a SMBH from $z =$ 6 - 20 \citep{smidt18} with Cloudy to determine out to what redshifts it could be found by {\em Euclid, WFIRST and JWST}.

\acknowledgments

We thank the anonymous referee for constructive comments that improved the quality of this paper and Gary Ferland and Peter van Hoof for very helpful discussions about Cloudy. DJW and SP were supported by the STFC New Applicant Grant ST/P000509/1 and ST/N504245/1 grant. TH is a JSPS International Research Fellow. EZ was funded by the Swedish National Space Board and SCOG was funded by the European Research Council via the ERC Advanced Grant STARLIGHT: Formation of the First Stars (project number 339177) and from the DFG via SFB 881 ``The Milky Way System" (sub-projects B1, B2, B8). AH was supported by an ARC Future Fellowship (FT120100363), by a Larkins Fellowship from Monash University, by TDLI though a grant from the Science and Technology Commission of Shanghai Municipality (Grants No.16DZ2260200) and National Natural Science Foundation of China (Grants No.11655002). LH was sponsored by the Swiss National Science Foundation (project number 200020-172505). Our simulations were performed on the Sciama cluster at the Institute of Cosmology and Gravitation at the University of Portsmouth.


\end{document}